\documentclass[aps,prl,twocolumn,superscriptaddress]{revtex4-1}
\usepackage{graphicx}
\usepackage{dcolumn}
\usepackage{bm}
\usepackage{natbib}
\usepackage{amsmath}
\usepackage{floatrow}

\begin{document}


\title{Imaging doubled shot noise in a Josephson scanning tunneling microscope}

\author{K.M. Bastiaans}
\affiliation{Leiden Institute of Physics, Leiden University, Niels Bohrweg 2, 2333 CA Leiden, The Netherlands}
\author{D. Cho}
\affiliation{Leiden Institute of Physics, Leiden University, Niels Bohrweg 2, 2333 CA Leiden, The Netherlands}
\author{D. Chatzopoulos}
\affiliation{Leiden Institute of Physics, Leiden University, Niels Bohrweg 2, 2333 CA Leiden, The Netherlands}
\author{M. Leeuwenhoek}
\affiliation{Leiden Institute of Physics, Leiden University, Niels Bohrweg 2, 2333 CA Leiden, The Netherlands}
\affiliation{Kavli Institute of Nanoscience, Delft University of Technology, Lorentzweg 1, 2628 CJ Delft, The Netherlands}
\author{C. Koks}
\affiliation{Leiden Institute of Physics, Leiden University, Niels Bohrweg 2, 2333 CA Leiden, The Netherlands}
\author{M.P. Allan}
\affiliation{Leiden Institute of Physics, Leiden University, Niels Bohrweg 2, 2333 CA Leiden, The Netherlands}
\email[]{allan@physics.leidenuniv.nl}


\begin{abstract}
We have imaged the current noise with atomic resolution in a Josephson scanning tunneling microscope with a Pb-Pb junction. By measuring the current noise as a function of applied bias, we reveal the change from single electron tunneling above the superconducting gap energy to double electron charge transfer below the gap energy when Andreev processes become dominant. Our spatially resolved noise maps show that this doubling occurs homogeneously on the surface, also on impurity locations, demonstrating that indeed the charge pairing is not influenced by disruptions in the superconductor smaller than the superconducting coherence length.
\end{abstract}



\maketitle

The coupling between two macroscopic superconducting electrodes through an insulating layer can lead to a dissipationless current called Josephson supercurrent. The critical current $I_{\text{C}}$ is the maximal supercurrent that the junction can sustain; it is related to the individual superconducting order parameters in both electrodes, as well as their coupling \cite{Josephson1963}. In the zero-voltage limit, this supercurrent is carried by paired electrons (Cooper pairs), carrying twice the electron charge $e$. Applying a bias voltage $V_{B}$ larger than twice the pair-breaking gap energy $\Delta$, $eV_{B} > 2 \Delta$, over the junction results in a normal current carried predominantly by quasiparticles with single electron charge (Fig. 1(b)). In the energy range below the gap edge, only so-called Andreev reflection processes can transport the quasiparticles across the junction by reflecting particles carrying opposite charge. These processes lead to effective charge transfer of multiple electron charges \cite{Blonder1982,Octavio1983,Beenakker1997}. In the energy range $\Delta < eV_{B} < 2 \Delta$, the dominant process is a single Andreev reflection leading to transfer of effectively double the electron charge, as illustrated in Fig. 1(c).

One cannot tell from the time-averaged value of the current whether it is carried by multiple integers of charge, but this becomes apparent when measuring the fluctuations of the current, or in others words, the current noise \cite{Blanter2000,Jehl2000,Lefloch2003,Ronen2016,Dieleman1997}. In general, the noise originating from the flow of uncorrelated particles in a tunneling junction (shot noise) is a purely Poissonian process. The current noise power $S_I$ is then proportional to the charge $q$ and the current $I$ of the carriers, $S=2q|I|$ \cite{Blanter2000}. At lower bias voltages, when Andreev processes become relevant, the transferred charge in a Josephson junction can effectively double and the noise is also expected to be two times the Poissonian value. Spectroscopic noise measurements in mesoscopic systems have revealed such noise signatures of multiple electron charge transport in superconducting junctions associated with Andreev processes \cite{Jehl2000,Lefloch2003,Ronen2016,Dieleman1997} and fractional charges in quantum hall systems \cite{de-Picciotto1997,Saminadayar1997}. 

In this Letter we perform such noise spectroscopy measurements spatially resolved with atomic resolution in a Josephson scanning tunneling microscope. We use our recently developed scanning tunneling noise microscopy (STNM) technique to spatially resolve the current and its time-resolved fluctuations simultaneously with atomic resolution \cite{Bastiaans2018}. We first demonstrate the current noise doubling from single to double charge transfer below the gap edge in a junction between a superconducting Pb tip and a Pb(111) sample. We then visualize this noise enhancement by spatially mapping the effective charge transfer over the sample surface. We show that it is homogeneous over the sample surface, also including impurity locations, demonstrating that the charge pairing is not influenced by disruptions smaller than the coherence length ($\xi \sim 80$ nm in Pb).

\begin{figure}
\includegraphics{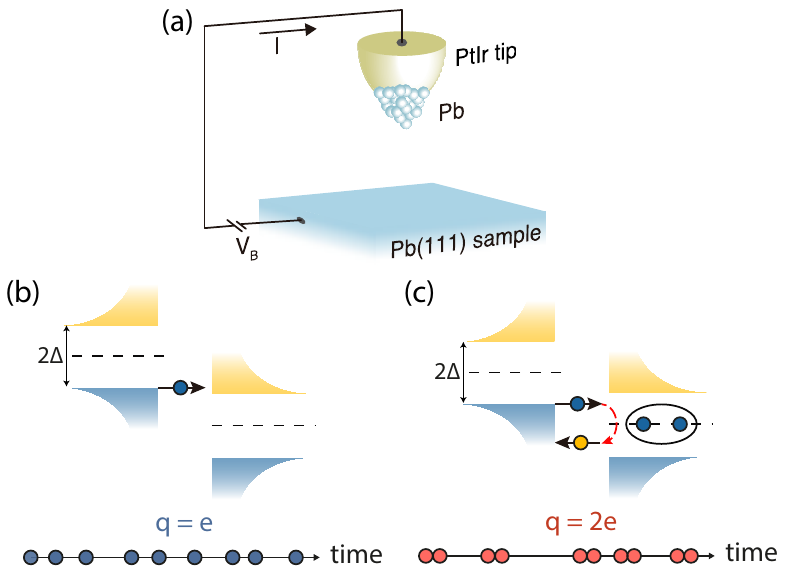}
\caption{(a) Schematics of a Josephson scanning tunneling microscope. The SIS junction consists of a Pb-coated tip ($\Delta_{\text{tip}} = 1.31$ meV) and a atomically flat Pb(111) surface ($\Delta_{\text{sample}} = 1.35$ meV) separated by a thin vacuum barrier. (b) Normal current carried by quasiparticles transferring single electron charge. The characteristic density of states of both superconducting electrodes are shown, with filled/empty states denoted by blue/yellow separated by the pair-breaking gap $2\Delta_{\text{tip/sample}}$. (c) Andreev reflection process. An electron transfers a Cooper-pair into the superconducting condensate by reflecting a hole in opposite direction, effectively transferring $2e$ charge.}\label{fig1}
\end{figure}

A schematic of our setup is shown in Fig. 1, where a superconducting STM tip is brought in tunneling contact with a superconducting sample to form a superconductor-insulator-superconductor (SIS) junction. We create this junction in our modified low-temperature (2.2 K) Unisoku USM-1500 STM setup. First, the Pb(111) single crystal surface is cleaned by repetitive cycles of $\text{Ar}^{+}$ sputtering at 1kV with an Ar pressure of $5.0 \times 10^{-5}$ mbar (background pressure $<1.0 \times 10^{-10}$ mbar) and annealing. We then indent the mechanically grinded PtIr tip into the surface to decorate it with a superconducting cluster of Pb atoms until we obtain a SIS junction \cite{Ruby2015,Franke2011,Randeria2016}. 

To demonstrate the high quality of the SIS junction in our setup, we display its distinct spectroscopic signatures for varying normal state resistance $R_{N}$ in Fig. 2. The first signature is visible in the single particle channel, where quasiparticles with energies larger than the pair-breaking gap transfer the charge. The tunneling spectra in Fig. 2(a) show sharp coherence peaks, which are located at energies equal to the sum of both superconducting gaps of tip and sample $\Delta = \Delta_{\text{tip}}+\Delta_{\text{sample}}=2.66$ meV. The clear U-shaped gap at 13.6 M$\Omega$ can be used as a benchmark for bulk-like superconducting properties of the tip and, also considering the low conductance, indicates that only a single transmission channel is present \cite{scheer1998}. Due to the sharp density of states of the superconducting tip, the spectroscopic features are much sharper than one would expect from conventional thermal broadening \cite{Ruby2015}. We can fit these spectra with a modified phenomenological gap equation \cite{Dynes1978} to extract the effective electron temperature of 2.2 K, which is similar to the measured phonon temperature, as electron-phonon coupling is still rather efficient at these temperatures. 

The next signature stems from Andreev processes that are visible at lower junction resistances. These lead to a sub-gap structure with humps in the differential conductance. Specifically, at energies below $2\Delta / n$, Andreev processes of order $n$ become possible with relative probability $\tau^{n+1}$,  where $\tau$ is the transparency of the junction. A small hump, indicated by the black arrows in Fig. 2 (a) and (b), is visible in the differential conductance when the order of lowest allowed Andreev reflection process changes \cite{Blonder1982,Octavio1983}. 

\begin{figure}
\includegraphics{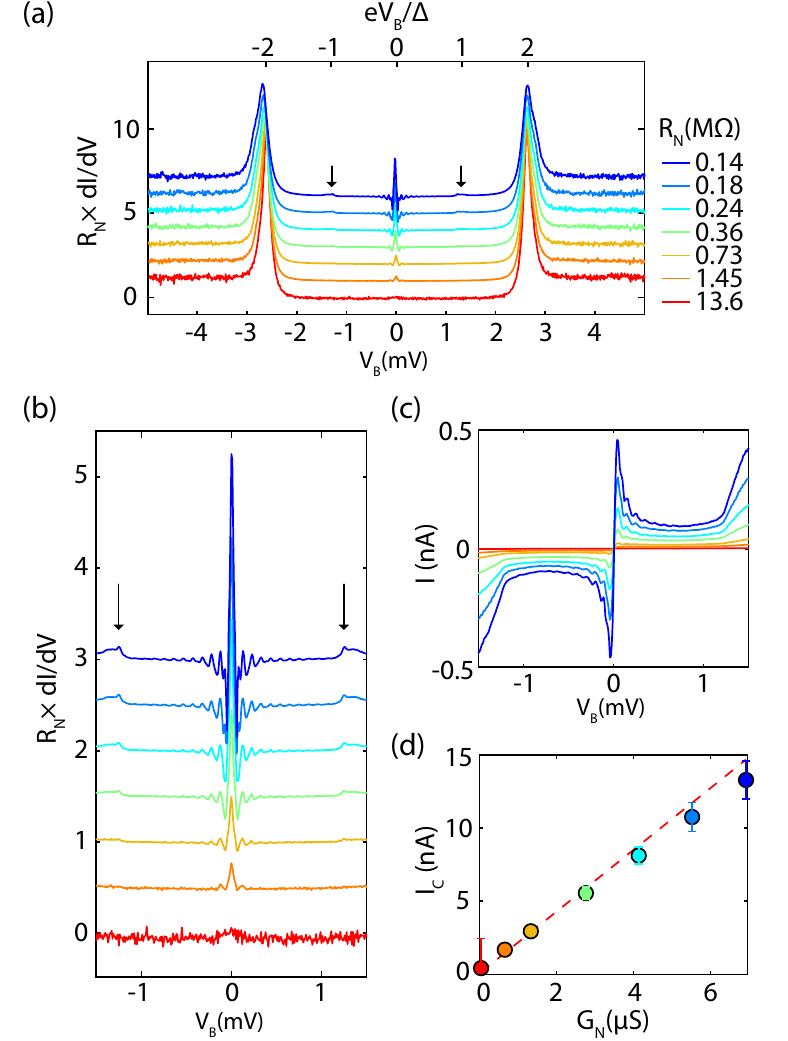}
\caption{(a) Differential conductance spectra multiplied by the normal state resistance acquired for variable setup conditions. Sharp coherence peaks can be seen at $\pm 2eV_{B}/ \Delta$. The increasing conductance around $\pm eV_{B}/ \Delta$ (arrows) with variable normal state resistance indicates the presence of Andreev processes. (b) Zoom in at the low energy features in the differential conductance spectra. The prominent peak at the Fermi energy that rises with decreasing $R_{N}$ is a signature of the Josephson supercurrent. (c) Current-voltage characteristics acquired simultaneously with the spectra shown in b. (d) Critical supercurrent of the junction (points) and its quantitative agreement with the Ambegaokar-Baratoff formula (red dashed line).}\label{fig2}
\end{figure}

The spectroscopic signatures related to the Josephson supercurrent in the junction are observed in the differential conductance at energies close to the Fermi level $E_F$: a peak that is enhanced with decreasing $R_N$, and small oscillations around the central peak (Fig. 2(b)). To understand these, we first survey the energy scales in our setup. The capacitive energy $E_C$ and the thermal energy are larger than the Josephson energy $E_J$. Therefore, the environmental impedance becomes a relevant quantity, and thermal phase fluctuations across the junction govern the Josephson current, shifting the maximum current to a non-zero bias \cite{Ingold1994}. Close to the Fermi energy $E_{F}$ we access the Cooper-pair channel associated with the coupling between the two superconducting condensates. The prominent peak at $E_{F}$ in Fig. 2(b), corresponding to the local maximum in the current in Fig. 2(c), originates from the phase fluctuating Josephson current \cite{Randeria2016,Ingold1994}. Both the maximum Josephson current and the differential conductance at zero bias are proportional to the square of the intrinsic critical current of the Josephson junction. The critical currents we obtain for these spectra show a linear relation with $R_N$ and are well consistent with the Ambegaokar-Baratoff formula (Fig. 2(d)) \cite{Ambegaokar1963}. We also note the small oscillating features in both the conductance and current spectra stretching far out to $\sim 1$ meV, originating from coupling of the junction with its dissipative electromagnetic environment, previously explained by a tip-induced antenna mode \cite{Randeria2016,Jack2015}. 

We now want to come to the central part of our paper, where we show the  visualization of  the doubling of the current noise in this scanning Josephson junction using  STNM. The central challenge for measuring current noise in a conventional scanning tunneling microscope (STM) is that the temporal resolution is generally limited to only a few kHz, because the combination of the high impedance ($\sim \text{G}\Omega$) tunnel junction and capacitance of the interfacing cables ($\sim 100$ pF) form an inherent low-pass filter. As a consequence, STM usually provides a static, time-averaged picture, lacking information about possible dynamical phenomena in the junction \cite{Chen2008}, especially when requiring atomic-resolution scanning. 

Our noise measurement apparatus, described in detail elsewhere \cite{Bastiaans2018}, builds upon earlier high-frequency STMs \cite{Birk1995,Kemiktarak2007} but is based on a superconducting LC resonating circuit that is connected to the Josephson junction, as illustrated in Fig. 3(a). Current fluctuations in the junction are converted into voltage fluctuations at resonance of the LC circuit, which are then amplified by the custom-built cryogenic amplifier \cite{Dong2014} into a 50 $\Omega$ line. To illustrate how we extract the magnitude of the current noise in the junction, we plot several curves of the measured power spectral density in Fig. 3(b) for various bias voltages. To mitigate the effect of a non-linear differential conductance on the effective resonator impedance, one needs to separate the measured signal into the noise components \cite{Bastiaans2018,Jabdaraghi2017}. The total measured voltage noise is
\begin{equation}
S_{\text{V}}^{\text{meas}} (\omega , V) =G^2 |Z_{\text{res}} |^2 S_I,
\end{equation}
where $G$ is the total gain of the amplification chain, $Z_{\text{res}}$ the impedance of the resonating circuit and $S_I$ the total current noise in the circuit. The strong influence of the highly non-linear differential conductance on the total impedance of the resonator is also illustrated in Fig. 3(b), where the clear change of the measured power spectral density for varying bias voltage is due to the simultaneously changing impedance of the resonator and amplitude of the noise as function of bias. 

After correcting for the non-linear differential conductance, the total current noise equals
\begin{equation}
S_{I}(I) = 2 q |I| \coth{\frac{qV}{2 k_{B} T}} + \frac{4 k_{B} T}{|Z_{\text{res}}|} + S_{\text{amp}},
\end{equation}
where $q$ is the effective charge, $T$ is the effective temperature and $S_{\text{amp}}$ is the input noise of the amplifier. The first term represents the shot noise in the junction and the second term represents the thermal noise. The black curves in Fig. 3(b) represent fitting of $S_{\text{V}}^{\text{meas}} (\omega, V)$ to the measured noise spectra, which we then use to obtain a value for the effective charge $q$ transferred across the junction.

\begin{figure}
\includegraphics{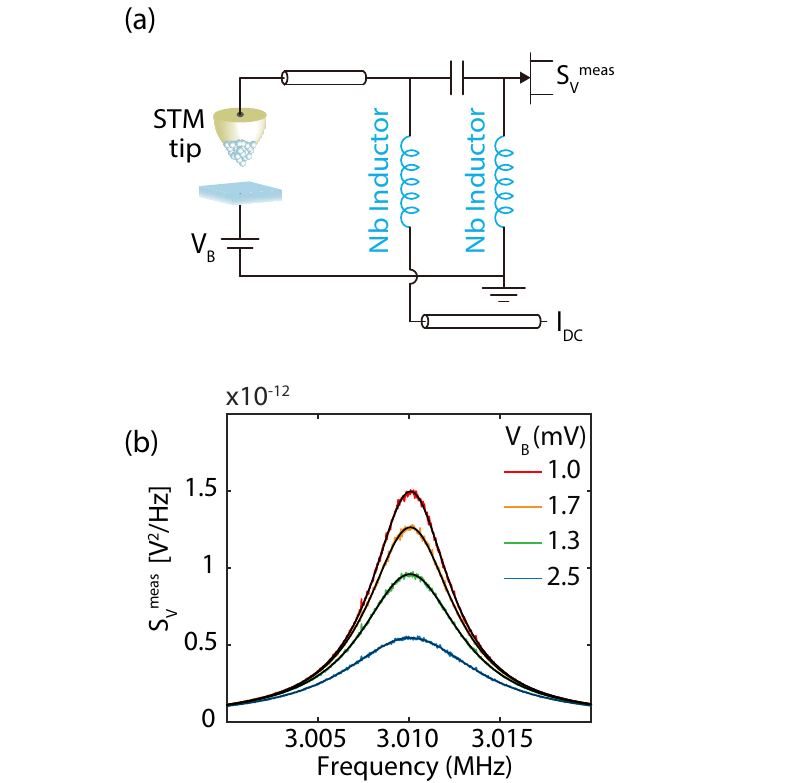}
\caption{(a) Measurement circuit of STNM, which allows for spatial mapping of the current fluctuations with atomic resolution. Superconducting niobium ($\text{T}_{\text{C}} = 9.2$ K, indicated in blue) inductors are used for the resonating circuit. (b) Power spectral density of the resonator circuit, in a small bandwidth around the resonance frequency. The different spectra represent various applied bias to the Josephson junction. Measured data is plotted by the colored lines, the black curves correspond to a circuit diagram fit.}\label{fig3}
\end{figure}

We first measure the current noise as a function of energy at a single location. Fig. 4(a) shows the measured current noise power as function of bias, with the zero-current noise subtracted to remove the thermal noise component and input noise of the amplifier: $S_{I}(I)-S_{I}(0)$. The dashed lines indicate the theoretically calculated shot noise curves as described above, for effective charge $e$ (blue) and $2e$ (red). At large bias voltage the experimental data follows the noise for single electron tunneling. However, the current noise clearly doubles from $e$ to $2e$ at the coherence peak energy $\pm 2eV_{B} / \Delta$. We obtain the effective charge transferred by dividing the measured noise power by the full Poissonian noise $S=2e|I|\coth{\frac{qV}{2 k_{B} T}}$ (Fig. 4(b)). The clear step in effective charge as a function of applied bias at the gap energy demonstrates that the tunneling current is now effectively carried by double charge quanta due to Andreev reflection processes. This is well consistent with theoretical descriptions \cite{Blonder1982,Octavio1983,Beenakker1997} and experimental observations in mesoscopic devices, where Andreev reflections lead to enhanced noise in nanofabricated SIS junctions \cite{Ronen2016,Dieleman1997} and short diffusive normal metal – superconductor contacts \cite{Jehl2000,Lefloch2003}, but has never been seen in a STM setup or at such low transparencies. In the present project, we use transparencies of $\tau \sim 10^{-3} - 10^{-4}$ leading to a single channel of transmission, whereas in mesoscopic devices usually multiple channels of $\tau \sim 10^{-1}$ are involved.
 
\begin{figure*}
\floatbox[{\capbeside\thisfloatsetup{capbesideposition={right,center},capbesidewidth=5cm}}]{figure}[\FBwidth]
{\caption{(a) Measured current noise for varying bias voltage acquired at a random location in Fig. 5(a). The dashed lines represent the calculated values for effective charge equals $e$ (blue) and $2e$ (red). (b) Effective charge transferred through the Josephson junction for varying bias. The data points represent $q=(S(I)-S(0))/2e|I|$, similar to the Fano factor. Dashed lines indicate $q=e$ and $q=2e$ lines. The black arrows indicate the bias voltage of the spatially resolved noise maps of Fig. 5.}\label{fig4}}
{\includegraphics{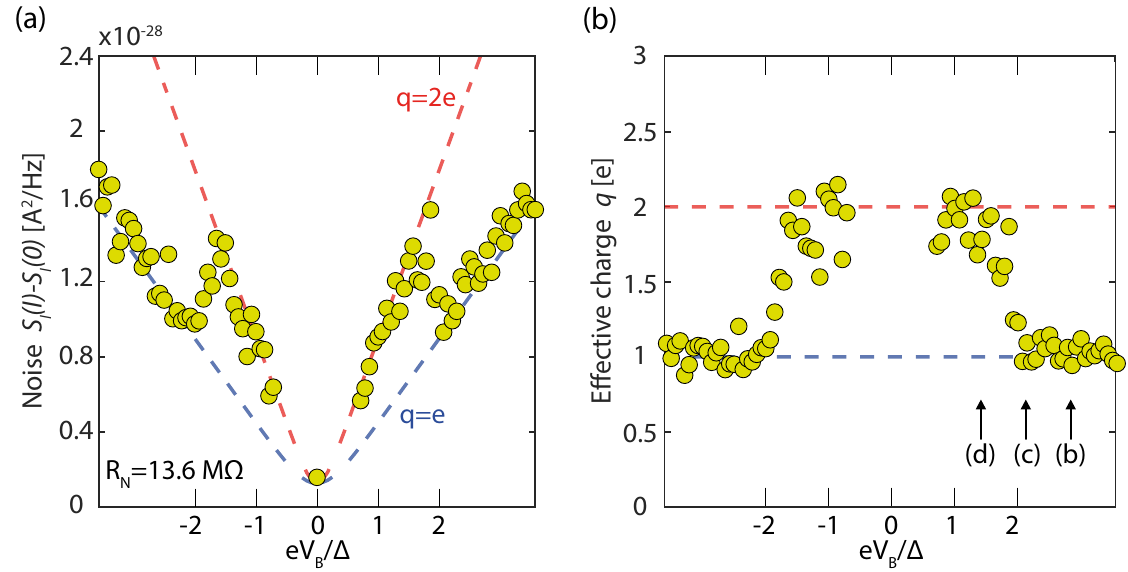}}
\end{figure*}

Finally, we apply the spatial mapping capabilities of our STNM setup to resolve this doubling of the noise over the sample surface with atomic-scale resolution. Figure 5(a) shows a topographic image of the Pb(111) surface in a 12,5 nm field of view, including a hexagonal shaped impurity previously identified as a sub-surface Ar nanocavity \cite{Muller2016}. We performed the noise-spectroscopy measurement in the same field of view by scanning the tip over the surface while simultaneously measuring the current noise. The spatially resolved noise maps at various bias voltages shown in Fig. 5 (b-d), exhibit a homogeneous contrast at energies above the pair-breaking gap energy, as is expected for transfer of uncorrelated particles. Below the superconducting gap energy we again observe homogeneous contrast, but now at an elevated value of the noise power around an effective charge equal to $2e$. While we observe a strong contrast in the topography, these spatially resolved noise maps show that the doubling occurs homogeneously over the surface, also on the location of the nanocavity. This demonstrates that disruptions in the superfluid, due to possible charge confinement or scattering on the nanocavity \cite{Muller2016}, on length scales smaller than the superconducting coherence length ($\xi \sim 80$ nm in Pb) do not influence the charge pairing, since the spatially resolved current noise is unaffected.

\begin{figure}
\includegraphics{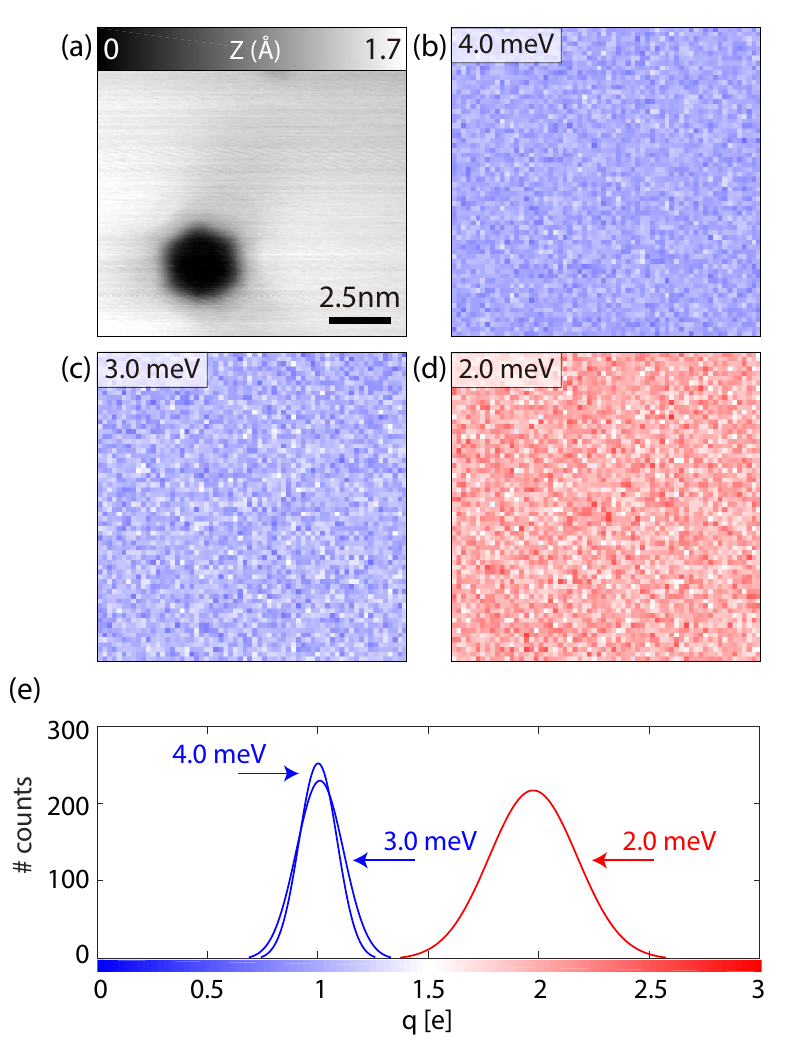}
\caption{(a) Topograph of 12,5 nm field of view on the Pb(111) surface, including a hexagonal shaped Ar nanocavity. (b-d) Spatial imaging of the effective charge for various bias voltages measured in feedback with a normal state resistance $R_{N} = 13.6 \text{ M}\Omega$.  The spatially resolved noise maps show homogeneous $q=e$ noise for $eV_{B} > 2\Delta$ (4.0 (b) and 3.0 (c) meV) and $q=2e$ noise for $eV_{B} < 2\Delta$ (2.0 meV (d)). (e) Histograms showing the distribution of the effective charge for each spatial noise map.}\label{fig5}
\end{figure}

In summary, we measured doubled shot noise caused by Andreev reflections in a Josephson scanning tunneling microscope using noise spectroscopy measurements. We spatially resolved this doubling with atomic-scale resolution on the surface of the conventional superconductor Pb(111). The ability to spatially resolve the charge dynamics with such precision opens new paths for investigating many-body correlation effects in quantum materials. Recently, it led to a novel understanding of cuprate high-temperature superconductors, where the discovery of charge trapping dynamics suggests a picture of copper-oxide planes separated by thin insulating layers within the three-dimensional superconducting state \cite{Bastiaans2018a,Massee2019}. Potentially, atomically resolved noise measurements will also reveal new insight in fluctuating stripe order \cite{Carlson2006} and pre-formed pairing in the pseudogap regime \cite{Lee2014,Keimer2015}, Kondo effects in heavy fermion systems \cite{Figgins2010}, or signatures of Majorana modes in one-dimensional wires on a superconducting surface \cite{Nadj-Perge2014}.

\begin{acknowledgments}
We acknowledge T. Benschop, Y.M. Blanter, V. Cheianov, T.M. Klapwijk, F. Massee, D.K. Morr, K. van Oosten, M.T. Randeria and J.M. van Ruitenbeek for valuable discussions. This work was supported by the European Research Council (ERC StG SpinMelt) and by the Netherlands Organization for Scientific Research (NWO/OCW), as part of the Frontiers of Nanoscience program, as well as through a Vidi grant (680-47-536). 
\end{acknowledgments}
%
\bibliography{Pb_noise_arXiv_bib}
\end{document}